\newcommand{\PRL}[3]{Phys.\ Rev.\ Lett.\ {\bf #1},\ #2 (#3)}
\newcommand{\RMP}[3]{Rev.\ Mod.\ Phys.\ {\bf #1},\ #2 (#3)}

\newcommand{\NAT}[3]{Nature\ {\bf #1},\ #2 (#3)}

\newcommand{\NATPHYS}[3]{Nature Phys.\ {\bf #1},\ #2 (#3)}

\newcommand{\PRA}[3]{Phys.\ Rev.\ A\ {\bf #1},\ #2 (#3)}
\newcommand{\PRB}[3]{Phys.\ Rev.\ B\ {\bf #1},\ #2 (#3)}

\newcommand{\JPB}[3]{J.\ Phys.\ B:\ At.\ Mol.\ Opt.\ Phys.\ {\bf #1},\ #2 (#3)}

\newcommand{\NJP}[3]{New\ J.\ Phys.\ {\bf #1},\ #2 (#3)}

\newcommand\x{\xi}

\newcommand{\diracslash}[1]{#1\llap{/\kern2pt}}

\newcommand{\be}{\begin{equation}}
\newcommand{\ee}{\end{equation}}
\newcommand{\bea}{\begin{eqnarray}}
\newcommand{\eea}{\end{eqnarray}}
\newcommand{\ba}[1]{\begin{array}{#1}}
\newcommand{\ea}{\end{array}}

\documentclass{ws-ijmpb}
\usepackage{graphicx}
\usepackage{verbatim}
\usepackage{color,hyperref}
\usepackage{amsfonts}
\definecolor{darkblue}{rgb}{0.0,0.0,0.7}
\definecolor{darkred}{rgb}{0.8,0.0,0.0}
\definecolor{darkgreen}{rgb}{0.0,0.8,0.0}
\hypersetup{colorlinks,breaklinks,
            linkcolor=darkred,urlcolor=darkblue,
            anchorcolor=darkblue,citecolor=darkgreen}

\begin{document}

\title{Effect of Weak Disorder on the BCS-BEC crossover in a two-dimensional  Fermi Gas}
\author{Ayan Khan}
\address{Department of Physics, School of Engineering and Applied Sciences,\\ Bennett University, Greater Noida-201310, India}
\author{B. Tanatar}
\address{Department of Physics, Bilkent University, 06800 Ankara, Turkey}


\date{\today}
\maketitle
\begin{history}
\received{Day Month Year}
\revised{Day Month Year}
\end{history}

\begin{abstract}
In this article we study the two-dimensional (2D) ultracold Fermi gas with weak impurity in the framework of mean-field theory where the impurity is introduced through Gaussian fluctuations. 
We have investigated the role of the impurity by studying the experimentally accessible quantities such as condensate fraction and equation of state of the ultracold systems.
Our analysis reveals that, at the crossover the disorder enhances superfluidity, which we attribute to the unique nature of the
unitary region and to the dimensional effect.    
 
\end{abstract}


\maketitle
\section{Introduction} 
The advances in experimental techniques in ultracold atom research
have widened the scope of our understanding of interacting quantum systems.
The early experiments and theoretical efforts on atomic gases made of bosonic particles\cite{stringari1} paved the way to intense research on fermionic systems \cite{regal}.
The observation of superfluidity in the Bardeen-Cooper-Schrieffer-Bose-Einstein condensate
(BCS-BEC) crossover regime has given impetus to the study of ultracold trapped Fermi gases at unitarity\cite{stringari2,dalibard1}.

Although the initial endeavor to understand the unitary Fermi gas was in the 3D systems, 
the interest lately is shifting towards lower dimensions.
It is well accepted that the ultracold atomic systems are good testing grounds for 
condensed matter physics and that provides an important motivation for exploring them in lower dimensions specially in 2D. At this juncture it must be noted that disorder in superconductor is a archetypal problem in condensed matter physics for over half a century. 
The seminal works of Anderson in the middle of the last century states that the weak disorder does not alter the superconducting critical temperature \cite{anderson} thereby paving the way for a intense study of superconductor-insulator transition. In two dimensions, the subject of superconductivity becomes highly intriguing as the existence of true long-range order is not possible.
Hence the BCS-BEC crossover in 2D receives great deal of interest for being the marginal dimension both for the classical fluctuation of the superfluid order parameter and formation of quantum bound states \cite{randeria1}.

Although the theoretical exploration of the BCS-BEC crossover in lower 
dimensional Fermi systems dates back a couple of decades \cite{randeria3,randeria4,sofo,quick},
in the last few years we have observed a new surge in research on 2D systems within the context of ultracold gases \cite{levinsen}. To note the superfluid behavior of a 2D Bose gas is demonstrated for $^{87}$Rb \cite{dalibard,campbell} and the pseudo-gap phenomena is observed in a 2D Fermi gas \cite{kohl,tempere}. The study of superfluidity and sound velocity within the mean field formalism has also been discussed \cite{salasnich}. Recently, a number of works discussed the connection between binding energy in 2D (usual controlling parameter) and 2D scattering length and extended their studies in the light of BCS-BEC crossover phenomenon\cite{giorgini,bertaina,werner,cao,zhang}. One can also find further extension to spin-orbit coupling and spin-mass imbalance \cite{chen,conduit}. A number of papers addressed the interplay of interaction and disorder effects in ultracold atomic systems \cite{palestini,tanatar,baur,yan,piraud,sherman,xian,beeler,krinner}. 

In this work, we study the 2D Fermi gas with weak disorder, where disorder is included through Gaussian fluctuations. Inclusion of fluctuations in lower dimensional ultracold Fermi gas is of current interest \cite{he} as this perturbative method opens the possibility of understanding the finite temperature properties \cite{big1,big2}. The basic model was proposed by Orso \cite{orso} for the 3D Fermi gas and later used to study different aspects of unitary Fermi gas \cite{khan,han,khan1}.  The 2D Fermi systems are usually characterized in terms of binding energy, although recently the 2D scattering length in analogy with its 3D counterpart has also started to be used. Thus,
in the preceding section we describe the connection between binding energy 
and 2D scattering length and the Gaussian fluctuation model which we use to incorporate the disorder. 
Afterwards we present our results for the pairing gap and chemical potential as well as the experimentally accessible quantities such as condensate fraction \cite{jochim1} and ground-state energy or equation of state \cite{jochim2} across the crossover. Finally, we draw our conclusion based on these results along with a short discussion on the possible experimental set-up of the disorder.

\section{The Model}
Since in 2D there exists a bound state ($\epsilon_{b}$) at any value of the attraction ($g$) unlike the 3D case,  $\epsilon_{b}$ 
is regarded as the traditional controlling parameter. It is also possible to use the 2D scattering length for this purpose in direct analogy with the 
3D systems\cite{bertaina}. For the case of an attractive square-well
the 2D scattering length can be written as $a=\rho_{0}\exp{\left[J_{0}(\kappa_{0})/\kappa_{0}J_{1}(\kappa_{0})\right]}$, where $\rho_{0}$ is the width of the well and $J_{0,1}$ are
the Bessel function. $\kappa_{0}=\sqrt{V_{0}m\rho_{0}^2}$ and $V_{0}$ defines the depth of the well. One can now connect the binding energy with the bare attractive potential via the 2D
scattering length through $\epsilon_{b}/E_{F}=-8/(ak_{F}e^{\gamma})^2$ \cite{randeria3} in terms of the Fermi wave vector $k_F$ and Fermi energy $E_F$, and $\gamma\simeq0.577$ is the Euler-Mascheroni's constant. 
For the attractive potential in 2D the scattering length is always non negative
and diverges at $V_0=0$ and at the zeros of $J_{1}$ corresponding to the appearance of
new two-body bound states in the well. Therefore, a bound state
is always present, however small the attraction is. Analogous to the situation in 3D 
the parameter $\xi=\ln{k_{F}a}>0$ (or $k_Fa>>1$) is considered as weak coupling regime (BCS) and
$\xi<0$ (or $k_Fa<<1$) is the strong coupling regime (BEC). Unitarity is defined at $\xi\thicksim0$ (or $k_Fa\thicksim1$). Equivalently, 
in terms of the binding energy $\epsilon_{b}$ from 2D scattering length we can consider $\epsilon_{b}/E_F<<2$ as BCS, $\epsilon_{b}/E_F>>2$ as BEC and the intermediate region as the crossover region. 

We are now in a position to discuss the case of weak disorder in the context of the mean-field model where we consider the inter-atomic interaction as short range (contact potential).
Thus the bound state (gap) equation
(after regularization) and density equation can be written as \cite{randeria3},
\begin{eqnarray}
\sum_{k}\left[\frac{1}{2\epsilon_{k}+\epsilon_{b}}-\frac{1}{2E_{k}}\right]=0,\,\,n=\sum_{k}\left[1-\frac{\xi_{k}}{E_{k}}\right],\label{2dgap}
\end{eqnarray}
where we use the usual BCS notation for single-particle energy 
$\xi_{k}=\mathbf{k}^2/2m-\mu$, quasi-particle energy $E_{k}=\sqrt{\xi_{k}^2+\Delta^2}$, and coherence factors $u_{k}^2=1/2(1+\xi_{k}/E_{k})$ and 
$v_{k}^2=1/2(1-\xi_{k}/E_{k})$.
The integrals can be evaluated analytically as shown by 
Randeria et al. \cite{randeria3}. 

The insertion of weak disorder has been carried out through Gaussian fluctuations.
For this purpose we assume that the range of the impurities should be much smaller 
than the average separation between them. 
To model it mathematically, we use the pseudo-potential as 
$\mathcal{V}_{d}(\mathbf{x})=
\sum_{i}g_{d}\delta(\mathbf{x}-\mathbf{x}_{i})$ 
where $g_{d}$ is the fermionic impurity coupling constant
(which is a function of impurity scattering length),
and $\mathbf{x}_{i}$ are the static positions of the quenched disorder. 
Thus, the correlation function is
$\langle \mathcal{V}_{d}(-q)\mathcal{V}_{d}(q)\rangle=
\beta\delta_{i\omega_{\nu},0}\gamma$
where $q=(\mathbf{q},i\omega_{\nu})$. 
$\beta=1/k_BT$ is the inverse temperature,
$\omega_{\nu}=2\pi \nu/\beta$ are the 
bosonic Matsubara frequencies with $\nu$ an integer. 
The disorder strength can be written as $\gamma=n_{i}g_{d}^2$, where
$n_{i}$ denotes the concentration of the impurities. 
Simple algebraic manipulations further reveal that $\gamma$ is a 
function of the relative size of the impurity.

The white noise is now inserted in the system by means of small fluctuations around the pairing gap as
$\delta\Delta(\mathbf{x},\tau)=\Delta(\mathbf{x},\tau)-\Delta$ where $\Delta$ is the homogeneous BCS pairing field.
The Green's function ($\mathcal{G}^{-1}$) can be written as a sum of Green's function in absence of disorder 
($G_{0}^{-1}=-\partial_{\tau}\mathbb{I}+(\nabla^2/2m+\mu)\mathbb{\sigma}_{z}+\Delta\mathbb{\sigma}_{x}$) 
and a self energy contribution 
($\Sigma=-\mathcal{V}_{d}\mathbb{\sigma}_{z}+\delta\Delta\mathbb{\sigma}_{+}+\bar{\delta\Delta}\mathbb{\sigma}_{-}$) 
which contains the disorder as well as the small fluctuations of the BCS pairing fields,
where $\sigma_{\pm}$ are the ladder matrices. 
The expansion of the inverse Nambu propagator ($\mathcal{G}^{-1}$) up to the second order 
it is possible to write the effective action ($\mathcal{S}_{eff}$)
as a sum of bosonic action ($\mathcal{S}_{B}$) 
and fermionic action ($\mathcal{S}_{F}$). 
There is an 
additional term which emerges from the linear order of self-energy 
expansion ($\mathcal{G}_{0}\Sigma$). It is possible to set the linear 
order term to zero, if we consider $\mathcal{S}_{F}$ is an extremum of 
$\mathcal{S}_{eff}$, after having performed all the fermionic 
Matsubara frequency sums. The modified thermodynamic potential
($\Omega$) can be written as a sum over fermionic ($\Omega_{F}$)
and bosonic ($\Omega_{B}$) potentials, 
$n=n_{F}+n_{B}=-\frac{\partial}{\partial\mu}(\Omega_{F}+\Omega_{B})=
-\frac{1}{\beta}\frac{\partial}{\partial\mu}(\mathcal{S}_{F}+\mathcal{S}_{B})$. 
It must be noted here that in the original mean-field description $\Omega_B=0$,
however here it contains the disorder contribution (as well as finite temperature contribution, which we neglect here as we are interested in impurity effect at zero temperature). Through $\Omega_F$ we yield the clean Fermi gas density equation (the first term in the right hand side of Eq.(\ref{dd})). Hence the manifestation of disorder in the
density equation (gap equation 
remains unaltered) appears as \cite{orso},
\begin{eqnarray}
n=\sum_{k}\left[1-\frac{\xi_{k}}{E_{k}}\right]+\frac{\partial\Omega_{d}}{\partial\mu},\label{dd}
\end{eqnarray}
where $\Omega_{d}$ is the disorder contribution in the bosonic thermodynamic potential which can be written as \cite{orso}
\begin{eqnarray}
\Omega_{d}=\frac{\gamma}{2}\sum_{q}\mathcal{N}^{\dagger}\mathcal{M}^{-1}\mathcal{N}.\label{thermo}
\end{eqnarray}
The inverse fluctuation propagator, $\mathcal{M}$, is a $2\times2$ symmetric matrix and at zero temperature it reads
\begin{eqnarray}
 \mathcal{M}_{11}&=&\frac{1}{g}+\sum_{k}\biggl[\frac{v_{k}^{2}v_{k+q}^{2}}{i\omega_{m}-E_{k}-E_{k+q}}-
\frac{u_{k}^{2}u_{k+q}}{i\omega_{m}+E_{k}+E_{k+q}}\biggr],\nonumber\\
\mathcal{M}_{12}&=&\sum_{k}u_{k}v_{k}u_{k+q}v_{k+q}\nonumber\\
&&\biggl[\frac{1}{i\omega_{m}+E_{k}+E_{k+q}}
-\frac{1}{i\omega_{m}-E_{k}-E_{k+q}}\biggr],\label{m}
\end{eqnarray}
with $\mathcal{M}_{22}(q)=\mathcal{M}_{11}(-q)$ and $\mathcal{M}_{21}(q)=\mathcal{M}_{12}(q)$. 
In Eq.(\ref{thermo}), $\mathcal{N}$ is a doublet which couples the disorder with the fluctuations. At $T=0$ after 
performing the fermionic Matsubara frequency summation one finds,
\begin{eqnarray}
 \mathcal{N}_{1}=\mathcal{N}_{2}=\sum_{k}\frac{\Delta(\xi_{k}+\xi_{k+q})}{2E_{k}E_{k+q}(E_{k}+E_{k+q})}.\label{fw}
\end{eqnarray}

Finally, the disorder induced bosonic thermodynamic potential at zero temperature can be written as \cite{khan2}
\begin{eqnarray}
 \Omega_{d}&=&\frac{\gamma}{2}\sum_{\mathbf{q}}\frac{2\mathcal{N}_{1}^2}{\mathcal{M}_{11}+\mathcal{M}_{12}}.\label{fo}
\end{eqnarray}
\begin{figure}
\begin{center}
\includegraphics[scale=0.35]{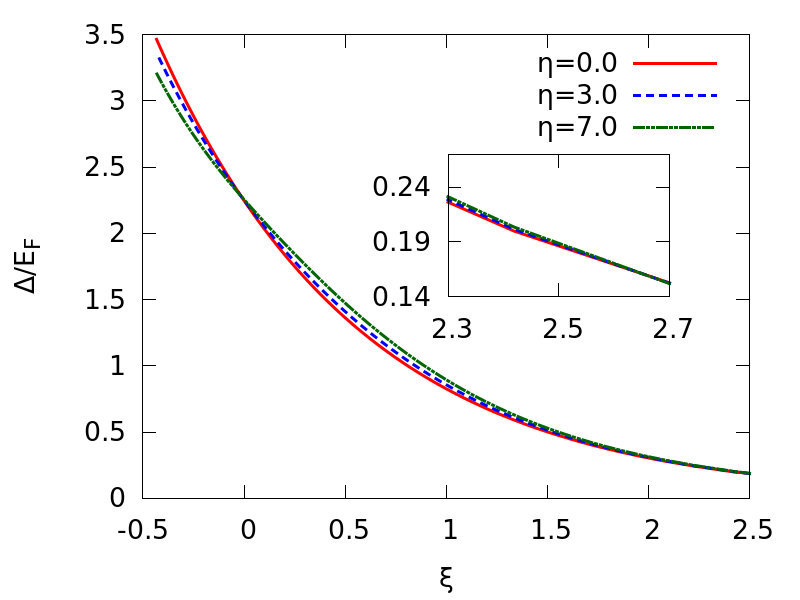}
\caption{Behavior of the order parameter as a function of interaction parameter 
$\xi$ and disorder parameter $\eta$. The solid line represent the clean Fermi gas ($\eta=0$) whereas 
the dashed line is for $\eta=3$ and dashed-dotted line is for $\eta=7$. In the inset, we zoom at the deep BCS regime where the bound state energy is very low.}\label{delta_sc}
\end{center}
\end{figure}

\section{Results and Discussion}
\subsection*{Pairing Gap and Chemical Potential}
The self-consistent numerical calculation of Eq.\,(\ref{dd}) along with the first equation of Eq.(\ref{2dgap}), yields the result for pairing gap which is depicted in Fig.\,\ref{delta_sc}. Here the disorder strength
is represented by $\eta$ which is a dimensionless quantity related to $\gamma$ i.e. $\eta=\gamma m^2/k_{F}=(3\pi^2/4)\gamma n/ \epsilon_{F}^2$.  
This implies that the impurity density and strength 
remains much less than the particle density $n$ and Fermi energy 
$\epsilon_{F}$.

As expected we observe a distinct depletion of paring gap (order parameter) in the BEC region ($\xi<0$) and in the BCS region ($\xi>0$) very little effect of disorder is observed (in the inset of Fig~\ref{delta_sc} we extend our calculation to further weak interaction regime to validate the conclusion). Similar features were obtained in
the 3D case \cite{khan} as well. However, here the behavior of $\Delta$ is different from its 3D counterpart in the crossover region ($\xi\in[0,1]$) where the pairing gap is enhanced due to disorder (as a function of $\eta$).
In this regime the attractive interaction is weak and as a consequence the particles are either moderately weakly bound 
or unbound, however the weak repulsive disorder potential pushes the unbound particles towards the attractive well and supports the pairing. Hence we observe an increase in the paring gap. 
In the strong coupling regime the pairs are strongly bound and disorder can only play a destructive role when it wins the competition with the inter-atomic attractive interaction
which results in depletion of pairing gap. Due to this interplay of disorder and interaction there exists a situation when interaction can completely mask the disorder effect and that happens just around $\xi\thicksim0$.
Hence  the impurity influenced $\Delta$ is unaffected in the deep BCS regime but it increases when we tune the binding energy from weak to strong. In the strong coupling BEC limit it again experiences depletion.
A precise calculation reveals that the pairing gap under the influence of impurity
matches with the clean limit around $\xi\thicksim-0.015$, exactly in the crossover region. 
This observation substantiates our surmise that unitary Fermi gas is less affected 
by disorder but rather supports superfluidity. The distinct enhancement of 
$\Delta$ around the crossover was not observed in the 3D system.

\begin{figure}
\begin{center}
\includegraphics[scale=0.35]{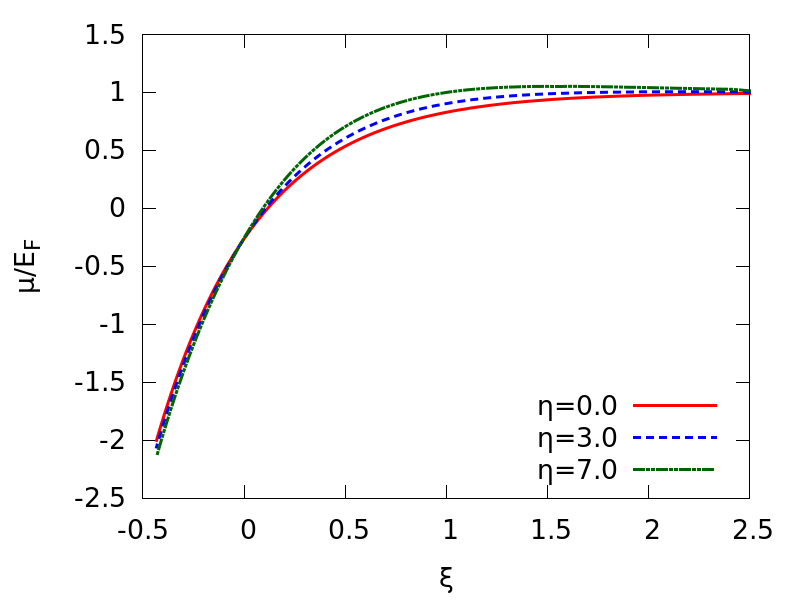}
\caption{Behavior of the chemical potential as a function of interaction parameter $\xi$ and disorder parameter $\eta$. The solid line represent the clean Fermi gas ($\eta=0$) whereas 
the dashed line is for $\eta=3$ and dashed-dotted line is for $\eta=7$.}\label{mu_sc}
\end{center}
\end{figure}
If we shift our attention to the chemical potential $\mu$ we observe a small change due to disorder and it is slightly higher in the crossover region and subsequently drops below the clean limit value (minutely) while reaching the strong coupling limit (described in Fig.~\ref{mu_sc}). 
Since bound pairs in the BEC limit are more prone to disorder 
we therefore observe lowering of chemical potential marginally in the strong coupling region where in the BCS limit the contribution through fluctuation is limited and the impurity effect is nominal.

\begin{figure}
\begin{center}
\includegraphics[scale=0.35]{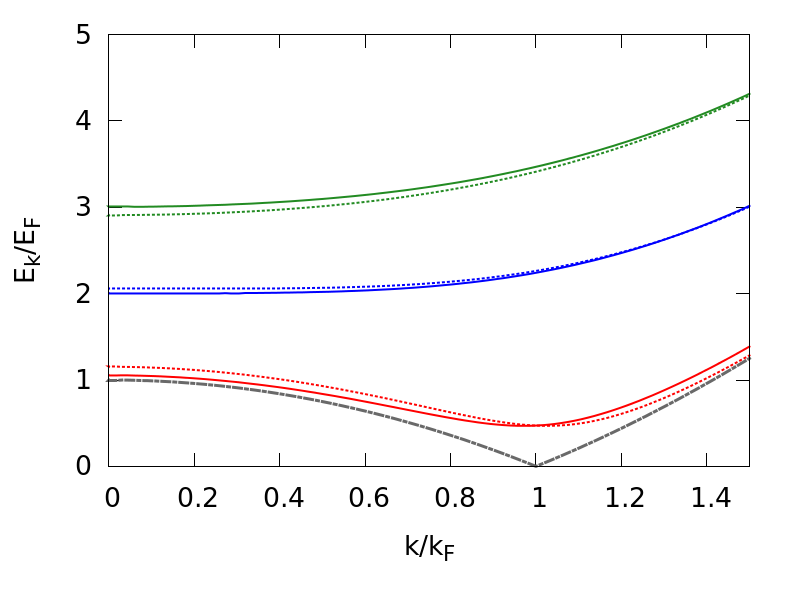}
\caption{Quasi particle dispersion as function of binding energy and disorder is depicted here. The solid line represent the clean Fermi gas ($\eta=0$) whereas 
the dashed line is for $\eta=7$. The lowest branch belongs to the BCS regime ($\xi\sim1.6$), where as the next higher branch represents the crossover region ($\xi\sim0$) and further higher branch depicts the BEC mode ($\x\sim-0.2$). The gray dashed dotted line is relates to the extremely weak coupling limit as $\epsilon_b\rightarrow0$ for clean Fermi gas.}\label{quasi}
\end{center}
\end{figure}

To probe the behavior of $\Delta$ and $\mu$ together as a function of disorder, we calculate the quasiparticle dispersion which is depicted in Fig.\,\ref{quasi}.
It is worth noting that the dispersion at the crossover and  BEC regime follows the Goldstone mode-like behavior whereas the BCS regime is more like roton dispersion. Figure\,\ref{quasi} also reveals
that the disorder effect is considerably low in the upper and lower branches and almost absent in the middle branch. The gray dashed-dotted line illustrates the clean limit dispersion for $\epsilon_b\rightarrow0$. In this limit the dispersion minimum occurs at $\sqrt{1-\epsilon_b/2}$ since we can replace $\Delta$ by $\sqrt{2\epsilon_b}$ and $\mu$ by $1-\epsilon_b/2$. Hence we can clearly see
that for $\epsilon_b\rightarrow0$, $\mu$ is pinned to $1$ and $E_k/E_F$ vanishes.  In the presence of disorder we cannot use the above mentioned analytic expressions for $\Delta$ and $\mu$. Therefore the excitation minimum never vanishes, but rather remains positive and finite ($E_k/E_F=\Delta$) albeit the minima occurs at $k/k_F\sim1$. Figure\,\ref{quasi} also suggests that the influence of disorder is more visible in the low momenta region.

To examine the unique behavior of negligible effect of disorder around the crossover (even enhancing the effect of superconducting pairing gap) more closely, we consider the long wavelength approximation. This approximation usually yields good results in the strong coupling (BEC) limit but it also provides a qualitative idea in the whole interaction range. 
Under this approximation the fluctuation propagation matrix and the doublet carrying the disorder effect can be expressed as \cite{taylor}
\begin{eqnarray}\label{M}
\mathcal{M}&=&\left(
\begin{array}{cc}
 \mathcal{A}+\mathcal{B} q^2 & \mathcal{A} \\
 \mathcal{A} & \mathcal{A}+\mathcal{B} q^2
\end{array}
\right),\quad
\mathcal{N}=\left(\begin{array}{c}\mathcal{N}_{1}\\\mathcal{N}_{2}\end{array}\right),
\end{eqnarray}
where $\mathcal{A}$ and $\mathcal{B}$ can be defined as \cite{strinati,taylor},
\begin{eqnarray}
\mathcal{A}&=&\sum_{k}\frac{\Delta^2}{4E_{k}^3}\nonumber\\
\mathcal{B}&=&\sum_{k}\frac{1}{32m}\left[\frac{\xi_{k}(2\xi_{k}^2-\Delta^2)}{E_{k}^5}+\frac{k^2}{\mathcal{2}m}\frac{\Delta^2(8\xi_{k}^2+3\Delta^2)}{E_{k}^7}\right]\, .\nonumber\\
\label{AB}
\end{eqnarray}
Also we consider at zero temperature the elements of the doublet $\mathcal{N}$ are equal i.e. 
$\mathcal{N}_{1}=\mathcal{N}_{2}$. Further, in the long wavelength strong coupling limit $q^2/2m<<\mu$ thus $\mathcal{N}_{1}$ can be written as \cite{orso}, 
\begin{eqnarray}
\mathcal{N}_{1}(q=0)=\sum_{k}\frac{\Delta\xi_{k}}{2E_{k}^3}\, .\label{W1}
\end{eqnarray}
To evaluate the integrals in Eqs.\,(\ref{AB}) and (\ref{W1}) we introduce the transformations $\mu/\Delta=x_{0}$ and $k^2/(2m\Delta)=x$. Thus, one can write 
$\xi_{k}=\xi_{x}/\Delta$,
$E_{k}=E_{x}/\Delta$ and $\xi_{x}=x-x_{0}$, $E_{x}=\sqrt{\xi_{x}^2+1}$. Now we can calculate the integrals in Eqs.\,(\ref{AB}) and (\ref{W1}) in a straightforward manner to obtain
\begin{eqnarray}\label{int}
\mathcal{N}_{1}&=&\frac{m}{4\pi\sqrt{1+x_{0}^2}},\,\,
\mathcal{A}=\frac{m \left(1+\frac{x_{0}}{\sqrt{1+x_{0}^2}}\right)}{8 \pi }\nonumber\\
\mathcal{B}&=&-\frac{1+x_{0} \left[4 \sqrt{1+x_{0}^2}+x_{0} \left\lbrace3+4 x_{0} \left(x_{0}+\sqrt{1+x_{0}^2}\right)\right\rbrace\right]}{48 \pi  \left(1+x_{0}^2\right)^{3/2} \Delta }\nonumber\\.
\end{eqnarray}
Now with the help of the above transformations and using some algebra we can rewrite Eq.\,(\ref{dd}) as
\begin{displaymath}
n=\frac{m\Delta}{2\pi}\left[x_{0}+\sqrt{1+x_{0}^2}\right]+3\eta\frac{k_{F}^2}{2\pi}I(x_{0},\epsilon_{b}),
\end{displaymath}
and 
\begin{equation}
\frac{\Delta}{E_{F}}=\frac{2\left[1+3\eta I(x_{0},\epsilon_{b}/E_F)\right]}{\left[x_{0}+\sqrt{1+x_{0}^2}\right]},\label{2dgapd}
\end{equation}
where $I$ is a logarithmic integral which is given as
\begin{eqnarray}
I(x_{0},\epsilon_{b}/E_F)&=&\frac{(f_{2}f_{3}-f_{1}f_{4})\frac{\epsilon_{b}}{E_F}}{f_{3}f_{4}(f_{3}\frac{\epsilon_{b}}{E_F}-6f_{4})}+\frac{f_{1}\ln\left|1-\frac{f_{3}\left(\epsilon_{b}/E_F\right)}{6f_{4}}\right|}{f_{3}^2},\nonumber\\\label{n_d}
\end{eqnarray}
where,
\begin{eqnarray}
f_{a}&=&\sqrt{1+x_{0}^2},\quad f_{b}=x_{0}+f_{a},\nonumber\\
f_{1}&=&\left(1+3x_{0}^2\right)\left[4f_{a}+x_{0}\left(5+4x_{0}f_{b}\right)\right],\quad f_{2}=f_{a}^2\left(1+2x_{0}f_{b}\right),\nonumber\\
f_{3}&=&1+x_{0}\left[4f_{a}+x_{0}\left(3 +4x_{0}f_{b}\right)\right],\quad f_{4}=f_{a}^2f_{b}.\nonumber
\end{eqnarray}

To estimate $\Delta$ and $\mu$ one requires numerics to solve Eq.\,(\ref{2dgapd}) and bound-state energy equation simultaneously. The bound-state equation after a change of variable can be written as \cite{randeria3,strinati}, 
\begin{eqnarray}
\frac{\epsilon_{b}}{E_{F}}&=&\frac{\Delta}{E_{F}}\left[\sqrt{1+x_{0}^2}-x_{0}\right].\label{eb1}
\end{eqnarray}
From Eq.\,(\ref{n_d}) we observe that the integral poses a branch cut and if we evaluate the branch point we obtain $\epsilon_b\thicksim1.9$ or $\xi\thicksim0.141$, which is very close to the the value where we observe an intersection between the clean and dirty values of the physical quantities.
Thus, almost zero effect of disorder in the vicinity of unitarity may not only be due to the unique inter-atomic correlation in the crossover region but also as a consequence of the dimensionality.
\subsection*{Condensate Fraction and Ground State Energy}
We now extend our analysis towards physically relevant quantities (experimentally observable) such as condensate fraction ($n_{0}$) and
ground-state energy ($E_0/(n\epsilon_F/2)$). For clean Fermi gas in 2D both of these quantities have been observed very recently \cite{jochim1,jochim2} in a clean system which makes our study relevant in the current context.
\begin{figure}
\begin{center}
\includegraphics[scale=0.35]{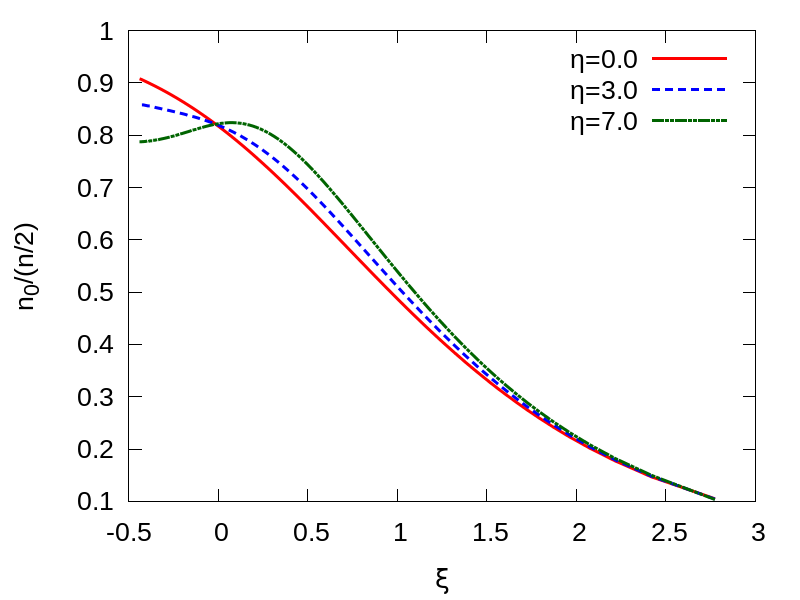}
\caption{Behavior of condensate fraction as a function of interaction parameter 
$\xi$ and disorder parameter $\eta$. The solid line depicts $\eta=0$ whereas $\eta=3$ and $7$ are represented by
for dashed and dashed-dotted lines, respectively. The figure illustrates a systematic depletion in the strong coupling regime and relative enhancement in the intermediate region.}\label{n0}
\end{center}
\end{figure}
The condensate fraction is calculated using the following formula \cite{manini}
\begin{eqnarray}
n_0&=&\sum_k\left[\frac{\Delta}{2E_k}\right]^2.
\end{eqnarray}
We observe in Fig.\,\ref{n0} that condensate fraction is higher near the crossover region, whereas in the BCS side there is no noticeable difference with the clean limit value. In the BEC side expectedly condensate fraction gets depleted. The behavior of condensate fraction is similar to the situation in 3D apart from the fact that the non-monotonic peak of the condensate fraction is higher than the clean limit result. In 3D, though we observed a non-monotonic behavior, the peak was always below the condensate fraction for a clean system. 

The ground-state energy can be written as \cite{stringari2}
\begin{equation}
E_{0}=\sum_{k}\left[\epsilon_{k}\left(1-\frac{\xi_{k}}{E_{k}}\right)-\frac{\Delta^2}{2E_{k}}\right].
\end{equation}
A suitable substitution of $k^2/(2m\Delta)=x$ and $\mu/\Delta=x_0$ leads to
\begin{eqnarray}
E_0&=&\frac{m\Delta^2}{2\pi}\int_{0}^{\infty}dx\left[x\left(1-\frac{\xi_{x}}{E_{x}}\right)-\frac{1}{2E_x}\right]\nonumber\\
&=&\frac{m\Delta^2}{8\pi}\left[2x_0\left(x_0+\sqrt{x_{0}^2+1}\right)-1\right]\nonumber\\
\frac{E_0}{n\left(\epsilon_F/2\right)}&=&\frac{\Delta^2}{\epsilon_F^2}\left[2x_0\left(x_0+\sqrt{x_{0}^2+1}\right)-1\right]\,
\end{eqnarray}
\begin{figure}
\begin{center}
\includegraphics[scale=0.35]{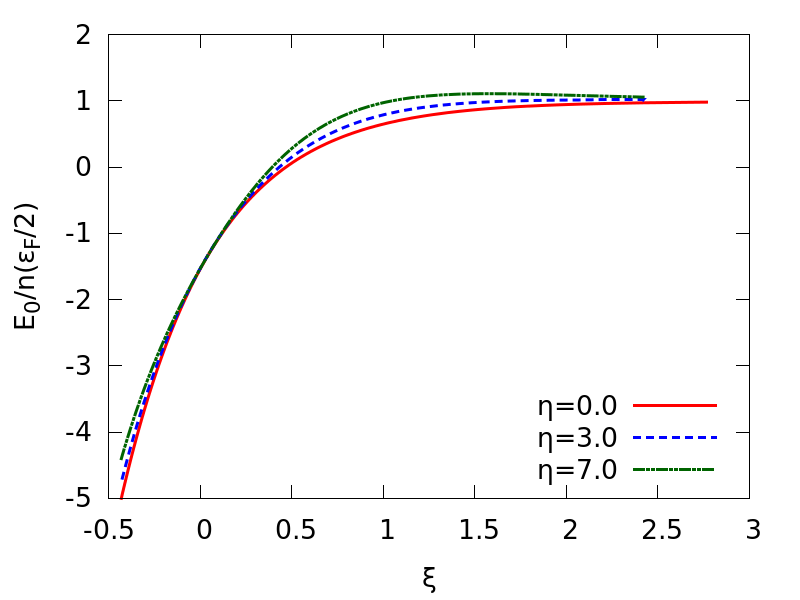}
\caption{The ground-state energy as a function of interaction parameter $\xi$ and disorder parameter $\eta$. The solid line represent the clean Fermi gas ($\eta=0$) whereas 
the dashed line is for $\eta=3$ and dashed-dotted line is for $\eta=7$.}\label{e0}
\end{center}
\end{figure}
where in the last line we divided the ground-state energy by number density $n$ to make the right-hand-side dimensionless.
Fig.\,\ref{e0} illustrates the behavior of the ground state energy ($E_0$) with weak disorder. We observe that $E_0$ is moderately higher in disordered environment in both weak and strong coupling limits. But in the intermediate region it converges to the clean limit value. 
This clearly implies that in the presence of quenched disorder crossover region is energetically most favorable. 
Expectedly, Fig.\,\ref{mu_sc} and \ref{e0} are very similar apart from their difference in normalization (as in 3D case). It is worth noting that the impurity induced ground-state energy is always higher than the clean limit value in both weak and strong coupling limits whereas $\mu$ drops below (the clean limit) in the BEC region. 

In this part we intend to discuss (i) how to realize the proposed disorder model experimentally and (ii) what we mean by weak disorder and consequently what range of values we can use.
In a recent experimental investigation it was shown that it is possible to tune the optical lattice depth in such a way that effectively one can observe the fermionic pairing mechanism from 
3D to 2D \cite{zweirlein,randeria2}. We borrow this idea and consider in analogy with the 3D system that randomly distributed light fermions are in an optical lattice whose depth is suitably modified
to shape as a 2D lattice and another optical trap (pancake in shape, whose longitudinal axis is heavily suppressed) in the same spatial dimension contains relatively heavy fermions. Because of
the disk-like shape of the potential, the degrees of freedom for the heavy fermions are only in the transverse direction. Moreover the optical trap frequencies are controlled in such a way that the lighter particles
cannot see the disk and the heavy fermions cannot see the 2D lattice. Hence they randomly collide with each other in the $x-y$ plane and the few heavy particles in this situation behave like quenched disorder. 

Now let us put forward the rationale behind the disorder parameter values used in the calculations and spare some thoughts on the acceptable values of disorder strength which can be considered as ``weak''. 
Previously in 3D, the weak value of $\eta$ ranges in the region $0\lesssim\eta\lesssim 0.7$, however here it is possible to further raise it up to $\eta\simeq 7$ (the normalization
of $\eta$ in both cases is $n_{F}/E_{F}^2$). The crucial point here is the dimensional difference which actually allows us the freedom to use higher $\eta$ values. If we consider the bare disorder strength is equal in both dimensions i.e. $\gamma_{2D}=\gamma_{3D}$ and we take the same density ($n_{F_{2D}}=n_{F_{3D}}$) then 
$\eta_{3D}/\eta_{2D}=E_{F_{(2D)}}^2/E_{F_{(3D)}}^2\simeq\pi^2/(3\pi^2)^{4/3}\simeq0.1$. This signifies that in 3D the admissible disorder strength will always be one order of magnitude less than the 2D counter part. Considering the elastic mean-free path for unbound fermions defined as 
$l_{F}=\pi/\gamma m^2$ in 3D, we then have $(k_{F}l_{F})_{3D}=4/(3\pi\eta_{3D})$. If we define the elastic mean-free path in 2D as $l_{F}=\pi k_{F}/\gamma m^2$, then $(k_{F}l_{F})_{2D}=2/\eta_{2D}$. Now if we consider that the mean-free path remains the same in 2D and 3D, then $\eta_{3D}\simeq0.2\eta_{2D}$. Therefore, the weak disorder limit in 2D is approximately one order of magnitude higher (in absolute numbers) than the 3D case.

\section{Conclusion}
In conclusion, we have explored the consequences of weak disorder in a 2D Fermi gas across the BCS-BEC crossover. We used the Gaussian fluctuation method to incorporate the impurity scattering and then solve the coupled equations self-consistently.
As a concise summary we conclude that,
(i) there is negligible effect of disorder on BCS side where the interaction is weak. (ii) The weak disorder seems to be supporting the superfluid nature in the crossover region as all physical quantities show enhancement.
(iii) Expectedly, the disorder starts destroying the superfluidity as we move to strong coupling region (BEC side) thereby 
pointing depletion in pairing gap and condensate fraction.
(iv) Since we observed a specific branch point in the disorder induced 
density contribution ($n_{d}=\partial\Omega_d/\partial\mu$) which has a logarithmic nature the effect of dimensionality might be attributed to the intersection of clean and dirty lines.

Keeping in mind the rapid progress in the experimental front we have also proposed a possible setup to experimentally realize the weak disorder effect in 2D ultracold Fermi systems.
Finally, we have included a short discussion on the acceptable values of disorder strength as ``weak''. We hope our analysis will motivate further work on this system especially  
studies based on Quantum Monte-Carlo simulations and Bogoliubov de-Gennes equation.   
  

\section*{Acknowledgement}
This work is supported by TUBITAK (112T176) and TUBA. AK likes to thank P. Pieri and L. Salasnich for their insightful comments.


\end{document}